\documentclass[epsf,12pt]{article}
\input psfig.sty
\usepackage{amsmath}
\include{amsfonts}
\newcommand{\beq}[2]{\begin{equation}#1\label{#2}\end{equation}}
\newcommand{\ceq}[1]{(\ref{#1})}
\newcommand{\mbd}[1]{\mbox{\bf #1}}

\newcommand{\bq}{\mbd{q}}

\newfont{\mbld}{cmbx10 scaled 800}
\newfont{\cab}{cmsy10 scaled 1200}
\newfont{\scab}{cmsy10 scaled 1000}

\begin{document}
\title{On A Bosonization Approach To Disordered Systems}
\author{Franco Ferrari\\
{\it Institute of Physics, University of Szczecin, ul. Wielkopolska 15,}\\
{\it 70-451 Szczecin, Poland}\thanks{e-mail:
fferrari@univ.szczecin.pl}.}


\maketitle

\abstract{
In \cite{sla} a new
bosonization procedure has been illustrated,
which allows to express a fermionic gaussian system in terms
of commuting variables at the price of
introducing an extra dimension. The  Fermi-Bose duality principle
established in this way  has many potential applications also
outside the context of gauge field theories in which
it has been developed. In this work we present an application to
the problem of averaging the correlation
functions with respect to random potentials in disordered systems
and similar problems. 
}
\vfill\eject
\pagestyle{plain}
\section{Introduction}
Both in statistical mechanics and in quantum mechanics there are
several  situations in which one has to average the
correlations functions of a physical system
with respect to disorder fields \cite{disosys}.
The averaging procedure is complicated by
the fact that the dependence of the correlation functions on the
disorder is not explicitly known. 
If we restrict ourselves to systems which admit a representation of
the Green functions in terms of gaussian fields,
one has two powerful tools at disposal
to solve this problem, the replica method \cite{repmet} and
the supersymmetric method \cite{supmet}.
The replica method can be applied to more general cases, but has 
the disadvantage that its mathematical consistency
has not yet been proved, so that
it has been subjected sometimes to some critiques concerning its
validity in nonperturbative calculations \cite{repcrit}.
On the other side, the introduction of supersymmetric fields requires a
fermionization of the system, in which the passage from
bosonic to fermionic degrees of freedom 
is often obscure. For this reason, difficulties arise
for instance in systems with spontaneous symmetry breaking,
because it is not easy to interprete the symmetry breaking in terms
of the resulting fermionic theory \cite{vilgis}.

In this letter we propose an alternative to the above two methods.
Our approach is very similar
in spirit to the supersymmetric one, but it involves only bosonic
fields. Instead of fermionization, we exploit the Fermi-Bose duality
principle introduced by A.~A. Slavnov
in the context of lattice gauge field
theories in order to provide an expression of
fermion determinants which does not contain anticommuting fields
\cite{sla}.
Successively, this principle  has been also applied in \cite{slaII}
to rewrite
the Faddeev-Popov determinants without anticommuting ghosts.
Since in our case all operators are hermitian,
it has been possible to simplify the
original procedure, which otherwise would have lead to
field theories with derivatives of the fourth order in the action.
Another relevant change with respect to \cite{sla, slaII} is the
choice of boundary conditions of the auxiliary bosonic fields.
Here boundary conditions are dictated by the compatibility requirement
with the
regularization needed
in the path integral approach to quantum mechanics
to guarantee convergence.

The material presented in this work is organized as follows.
In Section \ref{sec:2} the problem of averaging over the disorder fields
is briefly discussed. In Section \ref{sec:3} our alternative method
based on bosonization is presented. The conclusions are drawn in
Section \ref{sec:concl}.
\section{The Averaging Problem In Disor\-dered Sys\-tems}\label{sec:2}
Let $H$ be a local and hermitian Hamiltonian describing a
system with $D$ degrees of freedom $\bq=(q_1,\ldots,q_D)$.
Further, we suppose that $H$ depends on a set of random potentials
$\vec{\varphi}(\bq)= (\varphi_1(\bq),\varphi_2(\bq),\ldots)$ with a
given distribution $P(\vec{\varphi})$. 
Hamiltonians of this kind  
are widely applied in quantum mechanics \cite{disosys}.
For instance, choosing
$\vec{\varphi}=(\varphi_1)$, one obtains the energy operator of a disordered
system
\beq{H=H_0+\varphi_1}{dissys}
consisting of a fixed Hamiltonian $H_0$ and a random perturbation
$\varphi_1(\bq)$. If for example $\varphi_1$ is a source of Gaussian noise,
the general form of $P(\varphi_1)$ is given by:
\beq{P(\varphi_1)={\rm exp}\left\{-\int d^Dq d^Dq'
\varphi_1(\bq)K(\bq,\bq')\varphi_1(\bq')\right\}}{gaudis}
Analogously, with the same formalism
it is possible to discuss the motion of $n-$dimensional
particles
immersed in an electromagnetic field with components  $A_i$,
$i=1,\ldots,n$,
putting
$\vec{\varphi}=(A_1,\ldots,A_D)$ and taking as ``distribution''
\beq{P(A_1,\ldots,A_D)={\rm exp}\{-iS_{QED}\}}{weiqed}
where
\beq{S_{QED}=\frac 1{4g}
\int d^Dx F_{ij}^2
}{qedact}
is the usual action of quantum electrodynamics in $n$
dimensions and $F_{ij}=\partial_iA_j-\partial_jA_i$.

In the following, we denote the quantum average and the average over 
disorder fields
with the symbols $\langle\quad\rangle$ and
$(\quad)_{\vec{\varphi}}$ respectively.
With this notation, the advanced and retarded Green functions
of the Hamiltonian $H$
are given by:
\beq{G_E^\pm(\bq,\bq';\vec{\varphi})=\lim_{\epsilon\to
0^+}\left\langle\bq\left|\frac 1{E\pm
i\epsilon-H}\right|\bq'\right\rangle}{grefundef}
where $E+i\epsilon$ is a complex
 parameter
with an arbitrary small imaginary part $i\epsilon$.
Relevant information about the system may be obtained
computing averages over $\vec{\varphi}$
of products of the above Green functions:
\beq{\left(
G_E^\pm(\bq,\bq';\vec{\varphi})G_{E'}^\pm(\bq,\bq';\vec{\varphi})
\cdots\right)_{\vec{\varphi}}=
\int{\cal D}\vec{\varphi}P(
\vec{\varphi})G_E^\pm(\bq,\bq';\vec{\varphi})G_{E'}^\pm(\bq,\bq';\vec{\varphi})
\cdots}{aveprogrefun}
To this purpose,  it is often convenient to use a representation of
$G_E^\pm(\bq,\bq';V)$ in terms of complex scalar fields
$\phi,\bar\phi$:
\beq{G_E^\pm(\bq,\bq';\vec{\varphi}))=
\frac 1{i Z_\pm}\int {\cal D} \phi {\cal D}
\bar\phi
\phi(\bq)\bar\phi(\bq'){\rm exp}\left\{ \pm i\int
d^D\bq\bar\phi(E\pm i\epsilon -H)\phi\right\}}{fietherep}
$Z_\pm$ represents the partition function of the field theory:
\beq{Z_\pm=\int {\cal D} \phi {\cal D}
\bar\phi{\rm exp}\left\{ \pm i\int
d^D\bq\bar\phi(E\pm i\epsilon -H)\phi\right\}}{parfun}

It is easy to realize that, even in the simple case of a single noise source
with gaussian distribution as in Eq.~\ceq{gaudis},
it is difficult to integrate over the random potentials
in the right hand side of Eq.~\ceq{aveprogrefun}
due to the presence of
the factor $Z^{-1}_\pm$ 
in the definition
of $G^\pm(\bq,\bq',E|\vec{\varphi})$ (see Eq.\ceq{fietherep}).
In fact, the partition function $Z_\pm$ is a functional depending on
$\vec{\varphi}$ in a complicated way.
In the next Section, it will be shown how it is possible
to perform the average with respect to the random potentials
 without introducing replica fields or
fermionic degrees of freedom.
\section{The Bosonization Method}{\label{sec:3}}
First of all, we note that the class of problems under investigation has a
gaussian nature, as it is shown by the field theory representation of
Eq.~\ceq{fietherep}, in which only gaussian fields are involved.
Due to this fact, for our aims  it will be sufficient to consider
only the average of a single Green function. Let us
study for instance the following average:
\beq{\left\langle G_E^-(\bq,\bq';\vec{\varphi})\right\rangle_{\vec{\varphi}}=
\int {\cal
D}\vec{\varphi}P(\vec{\varphi})G^-_E(\bq,\bq';\vec{\varphi})}{sinfunave}
Now it will be convenient to interpret the factor $Z^{-1}_\pm$ in
Eq.~\ceq{fietherep} as the functional determinant of the operator
$E-H$:
\beq{Z^{-1}_\pm={\rm det}(E\pm i\epsilon-H)}{detrep}
To express the determinant appearing in the right hand side of
Eq.~\ceq{detrep}, we apply the Fermi-Bose duality principle
proposed in \cite{sla}.
To this purpose, we introduce a fictitious time $\tau$ such
that
\beq{-T\le\tau\le T}{fictimran}
and two sets of
auxiliary fields $c_n(\bq,\tau),\bar c_n(\bq,\tau)$ and $\chi_n(\bq)$,
$n-1,2$. The field
$\bar c_n$ is the hermitian conjugate of $c_n$
\beq{\bar c_n
=(c_n)^\dagger
}{hercon}
while $\chi_n$ is
an hermitian scalar field, i.e. $(\chi_n)^\dagger=\chi_n$.
The fields $c_n$ and $\bar c_n$ satisfy the boundary conditions:
\beq{c_n(\bq,-T)=B_n(\bq)\qquad\qquad\qquad
\bar c_n(\bq,T)=B_n^\dagger(\bq)}{boucon}
where $B_n(\bq)$ is an arbitrary function of $\bq$.

We are now ready to prove the following formula:
\beq{Z_-^{-1}={\rm det}(E+i\epsilon-H)=\lim_{T\to +\infty}
{\cal Z}_{c,1}{\cal Z}_{c,2}
}{baside}
where
\beq{{\cal Z}_{c,n}=
\int{\cal D}c_n{\cal D}\bar c_n{\cal D}\chi_n e^{iS_n}}
{zcparfun}
and
\begin{eqnarray}
\lefteqn{S_n=\int_{-T}^Td\tau\int d^Dq\left[-\left(
\frac i2\frac{\partial\bar c_n}{\partial\tau}+(E-H)\bar c_n\right)c_n\right.}
\nonumber\\
&&+\left.
\left(
\frac i2\frac{\partial c_n}{\partial\tau}-(E-H)c_n\right)\bar c_n
+2i\epsilon c_n\bar c_n+
\chi(\bar c_n +c_n)\right]\label{scact}
\end{eqnarray}
The proof goes as follows.
Since the operator $E-H$ is hermitian, it supports a complete system of
orthonormal  eigenfunctions $\psi_\alpha$ with eigenvalues $\lambda_\alpha$:
\beq{(E-H)\psi_\alpha(\bq)=\lambda_\alpha\psi_\alpha(\bq)}{eigfundef}
Thus, it is possible to expand the fields $c_n,\bar c_n$ and
 $\chi_n$ in terms of
the $\psi_\alpha$:
\begin{eqnarray}
c_n(\bq,\tau)&=&\sum_\alpha c^\alpha_n(\tau)\psi_\alpha(\bq)
\label{aeigfunexp}\\
\bar c_n(\bq,\tau)&=&\sum_\alpha \bar c^\alpha_n(\tau)\psi_\alpha(\bq)
\label{beigfunexp}\\
\chi_n(\bq)&=&\sum_\alpha\chi^\alpha_n\psi_\alpha(\bq)\label{ceigfunexp}
\end{eqnarray}
Here $c^\alpha_n(\tau),\bar c^\alpha_n(\tau)$ depend only on the pseudo-time
$\tau$, while the $\chi^\alpha_n$ are constant coefficients.
To these equations, one should add the expansions of the fields
$B_n$ and $B^\dagger_n$ which express the boundary conditions:
\beq{B_n(\bq)=\sum_\alpha B^\alpha_n\psi_\alpha(\bq)
\qquad\qquad\qquad
B^\dagger_n(\bq)=\sum_\alpha \bar B^\alpha_n\psi_\alpha(\bq)}{bocoexp}
Substituting Eqs.~(\ref{aeigfunexp}--\ref{ceigfunexp})
in the action \ceq{scact} and remembering that:
\beq{\int d^Dq\psi_\alpha(\bq)\psi_\beta(\bq)=\delta_{\alpha\beta}}
{ortnordef}
one obtains for ${\cal Z}_{c,n}$:
\beq{{\cal Z}_{c,n}=\prod_\alpha\int{\cal D}\bar c^\alpha_n
{\cal D}c^\alpha_n{\cal D}\chi^\alpha_n e^{iS_{n,\alpha}}}{actaftnmexp}
with
\begin{eqnarray}
\lefteqn{S_{n,\alpha}=\int_{-T}^Td\tau
	\left[
		-\left(
			\frac i2\frac{\partial\bar c^\alpha_n}
			{\partial\tau}
			+\lambda_\alpha\bar c^\alpha_n
		\right)
	c^\alpha_n\right.}\nonumber\\
&&+\left.\left(\frac i2\frac{\partial c^\alpha_n}{\partial\tau}
		-\lambda_\alpha c^\alpha_n
		\right)
		\bar c^\alpha_n+2i\epsilon c^\alpha_n\bar c^\alpha_n
+\chi^\alpha_n(\bar c^\alpha_n+c^\alpha_n)\right]\label{essealpha}
\end{eqnarray}
The gaussian path integral over the fields $\bar c^\alpha_n$ and
$c^\alpha_n$ may be easily computed with the saddle point method. To
this purpose,  one has to solve the classical equations of motion of
these fields:
\begin{eqnarray}
\dot c^\alpha_n+\omega_\alpha c^\alpha_n-i\chi^\alpha_n&=&0\label{1equmot}\\
\dot{\bar c}^\alpha_n-\omega_\alpha c^\alpha_n+i\chi^\alpha_n&=&0
\label{2equmot}
\end{eqnarray}
In the above equation we have put $\dot c^\alpha_n=dc^\alpha_n/d\tau$,
$\dot{\bar c}^\alpha_n=d\bar c^\alpha_n/d\tau$ and
\beq{\omega_\alpha=2(i\lambda_\alpha+\epsilon)}{omedef}
The solutions of (\ref{1equmot}--\ref{2equmot}) satisfying the desired boundary
conditions are:
\begin{eqnarray}
c_{n, cl}^\alpha(\tau)&=&e^{-\omega_\alpha(\tau+T)}B^\alpha_n+
i\frac{\chi^\alpha_n}
{\omega_\alpha}\left[1-e^{-\omega_\alpha(\tau+T)}\right]
\label{clasol1}\\
\bar c_{n,cl}^\alpha(\tau)&=&e^{\omega_\alpha(\tau-T)}\bar B^\alpha_n+
i\frac{\chi^\alpha_n}
{\omega_\alpha}\left[1-e^{\omega_\alpha(\tau-T)}\right]
\label{clasol2}
\end{eqnarray}
Let us note that $c_{n,cl}^\alpha$ and $\bar c_{n,cl}^\alpha$ do not
diverge for large values of $\tau$. Moreover, it is clear that the
boundary conditions are irrelevant in the limit
$T\rightarrow+\infty$, because their contribution vanishes exponentially
as $e^{-2\epsilon T}$.

After the field transformation:
\begin{eqnarray}
c^\alpha_n(\tau)=c^\alpha_{n,cl}(\tau)+c^\alpha_{n,q}(\tau)\label{1tra}\\
\bar c^\alpha_n(\tau)=
\bar c^\alpha_{n,cl}(\tau)+\bar c^\alpha_{n,q}(\tau)\label{2tra}
\end{eqnarray}
${\cal Z}_{c,n}$ becomes:
\beq{{\cal Z}_{c,n}=\lim_{T\to+\infty}\prod_\alpha\int{\cal D}\chi^\alpha_n
{\cal N}_{n,\alpha}
e^{iS_{n,\alpha}^{cl}}
}{zcnew}
Here we have put
\begin{eqnarray}
\lefteqn{S_{n,\alpha}^{cl}=\frac 12\int_{-T}^Td\tau\left[
2i\frac{(\chi^\alpha_n)^2}
{\omega_\alpha}-i\frac{(\chi^\alpha_n)^2}{\omega_\alpha}
e^{-\omega_\alpha T}\left( e^{-\omega_\alpha
\tau}+e^{\omega_\alpha\tau} \right)\right.}\nonumber\\
&&\left.\phantom{\frac{(\chi^\alpha_n)^2}{\omega_\alpha}}\!\!\!\!\!\!\!\!\!\!
\!\!\!\!\!\!\!\!\!+
 i\chi^\alpha_n e^{-\omega_\alpha
T}\left(e^{-\omega_\alpha\tau}B^\alpha_n
+e^{\omega_\alpha\tau}\bar B^\alpha_n\right)\right]\label{scl}
\end{eqnarray}
and
\beq{
{\cal N}_{n,\alpha}=\int{\cal D}c^\alpha_{n,q}{\cal D}\bar c^\alpha_{n,q}
\exp\left[i\int_{-T}^T\left(-\frac i2\dot{\bar c}^\alpha_{n,q}
c^\alpha_{n,q}+
\frac i2\dot c^\alpha_{n,q}\bar c^\alpha_{n,q}\right)\right]}{qint}
It is easy to show that the constant factor ${\cal N}_{n,\alpha}$
produced by the integration over the ``quantum'' fields $c_{n,q}^\alpha,\bar
c_{n,q}^\alpha$ is a just a constant, which is independent of
$\omega_\alpha$ and thus can be ignored\footnote{Let us note that
this factor is also independent on $T$.}.
At this point it is possible
to perform the integration over the pseudo-time $\tau$ in the action
$S_{n,\alpha}^{cl}$. The result is:
\beq{
S_{n,\alpha}^{cl}=2i\frac{(\chi^\alpha_n)^2T}{\omega_\alpha}
-i\frac{(\chi^\alpha_n)^2}{\omega_\alpha^2}\left(
1-e^{-2\omega_\alpha T}\right)+
\frac{\chi^\alpha_n}{\omega_\alpha}\left(1-e^{-2\omega_\alpha T}\right)
(B^\alpha_n+\bar B^\alpha_n)}{actaftintpt}
Finally, one has to integrate over the variables $\chi^\alpha_n$
in ${\cal Z}_{c,n}$:
\begin{eqnarray}
\lefteqn{
{\cal Z}_{c,n}
=}\nonumber\\
&&\int{\cal D}\chi^\alpha_n
\exp\left\{i
\left[
2i
\frac{(\chi^\alpha_n)^2T}
{\omega_\alpha}
-i\frac{(\chi^\alpha_n)^2}
{\omega_\alpha^2}
\left(
1-e^{-2\omega_\alpha T}
\right)
\right.
\right.
\nonumber\\
&&\left.\left.
\phantom{\frac{(\chi^\alpha_n)^2}{\omega_\alpha}}\!\!\!\!\!\!\!\!\!\!
\!\!\!\!\!\!\!\!\!+
\frac{\chi^\alpha_n}{\omega_\alpha}\left(1-e^{-2\omega_\alpha T}\right)
(B^\alpha_n+\bar B^\alpha_n)\right]\right\}\label{zcapreoi}
\end{eqnarray}
Only the first term in the right hand side of the above equation
becomes relevant when $T$ becomes very large, as it is in our case.
Since
\beq{\frac{2iT}{\omega_\alpha}=T\frac{\lambda_\alpha +i\epsilon }
{\lambda_\alpha^2+\epsilon^2}
}{convcond}
it turns out that, thanks to the presence of
the $\epsilon-$term in the action $S_n$ of Eq.~\ceq{scact},
the integrals in ${\cal D}\chi^\alpha_n$
are convergent.
Upon renormalizing the fields $\chi^\alpha_n$ in Eq.~\ceq{zcapreoi}
as follows:
$\chi^{\alpha\prime}_n=\chi^\alpha_n T^{1/2}$, one finds:
\beq{{\cal Z}_{c,n}=\prod_\alpha\sqrt{\lambda_\alpha+i\epsilon}
=\sqrt{\det(E-H+i\epsilon)}}{finres}
This proves Eq.~\ceq{baside} as desired.
An analogous formula can be derived for $Z_+^{-1}$.

Coming back to the original averaging problem, we rewrite
Eq.~\ceq{sinfunave} in the form:
\begin{eqnarray}
\lefteqn{\left\langle G_E^-(\bq,\bq';\vec{\varphi})\right
\rangle_{\vec{\varphi}}=
\lim_{T\to+\infty}\int {\cal
D}\vec{\varphi}P(\vec{\varphi})
{\cal Z}_{c,1}{\cal Z}_{c,2}}\nonumber\\
&\times&\int{\cal D}\phi{\cal D}\bar\phi
\phi(\bq)\bar\phi(\bq')\exp\left\{-i\int_{-T}^T\frac{d\tau}{2T}\int d^D\bq
\bar \phi(E-i\epsilon-H)\phi\right\}\label{finforfunave}
\end{eqnarray}
The dependence of $G_E^-(\bq,\bq';\vec{\varphi})$ on the
disorder fields $\vec\varphi$ is now
explicit and it is given by Eq.~\ceq{scact}, which expresses the inverse of the
partition function $Z_-$ as a path integral over gaussian fields
$c_n,\bar c_n,\chi_n$, $n=1,2$. At this point it is possible to perform
the averaging over the disorder fields at least
perturbatively. In the case of Hamiltonians
like that of Eq.~\ceq{dissys},
the random potential $\varphi_1$
 may be integrated out from the partition function
with the help of a gaussian integral.

Let us note that in Eq.~\ceq{finforfunave} the limit for $T\rightarrow+\infty$
has been permuted with the integration over the disorder fields $\vec \varphi$.
In a similar way, in the method of replicas one needs to permute
the limit of vanishing replica's and the average with respect to
the disorder. The difference between the two approaches
is that in the present case the limit
$T\rightarrow +\infty$ does not require a complex analytical
continuation
as in the method of replicas and it is mathematically under control.
In fact, the presence of the variable $T$ is only limited to the
partition
functions ${\cal Z}_{c,n}$ of the bosonic fields $c_n,\bar
c_n,\chi_n$.
These partition functions may always be rewritten as in
Eq.~\ceq{zcapreoi}, i.~e.
in terms of standard integrals over the real variables
$\chi^\alpha_n$, which are convergent in the limit $T\rightarrow
+\infty$ due to the presence of the $\epsilon$ term.
As a consequence, the permutation of the symbol $\lim_{T\to +\infty}$
with the integrals which are necessary to compute ${\cal Z}_{c,n}$
is allowed.
\section{Conclusions}\label{sec:concl}
Concluding, in this work it has been presented
a bosonization approach to
disordered systems based on the
Fermi--Bose principle of \cite{sla}, which is alternative to
the supersymmetric method and to the method of
replicas.
The bosonization
approach does not contain fermionic degrees of freedom because
it involves only bosonic fields. As a consequence,
it may be useful in the investigations of
systems with spontaneous symmetry breaking, which are sometimes 
complicated by the presence
of anticommuting variables.

Let us note that in the case of quantum chromodynamics
the Fermi-Bose principle
used to express the Faddeev--Popov determinant without the help
of ghost fields leads
to a new symmetry, which replaces the usual BRST symmetry.
It would be thus
interesting to check if also the disordered path integral
of Eq.~\ceq{finforfunave} enjoys an analogous new symmetry,
which would  replace
the fermion-boson symmetry of the supersymmetric method.

\end{document}